\documentclass[sn-mathphys,Numbered]{sn-jnl}

\renewenvironment{table}[1][]%
{\tableorg[#1]%
\tablebodyfont%
\renewcommand\footnotetext[2][]{{\removelastskip\vskip3pt%
\let\tablebodyfont\tablefootnotefont%
\hskip0pt\if!##1!\else{\smash{$^{##1}$}}\fi##2\par}}%
}{\endtableorg}

\usepackage{graphicx}%
\usepackage{multirow}%
\usepackage{amsmath,amssymb,amsfonts}%
\usepackage{amsthm}%
\usepackage{mathrsfs}%
\usepackage[title]{appendix}%
\usepackage{textcomp}%
\usepackage{manyfoot}%
\usepackage{booktabs}%
\usepackage{algorithm}%
\usepackage{algorithmicx}%
\usepackage{algpseudocode}%
\usepackage{listings}%
\usepackage[table]{xcolor}%
\usepackage{caption}%
\usepackage{float}%
\usepackage{geometry}%
\usepackage{graphicx}%
\usepackage{multirow}%
\usepackage{colortbl}%
\usepackage{subcaption}%
\usepackage{xurl} %

\definecolor{lightorange}{RGB}{255,200,150} 
\definecolor{redpink}{RGB}{255,105,97} 
\definecolor{green}{RGB}{144,238,144} 

\theoremstyle{thmstyleone}%

\theoremstyle{thmstyletwo}%

\theoremstyle{thmstylethree}%

\begin{document}

\title[Phishing Website Detection through Multi-Model Analysis of HTML Content]{Phishing Website Detection through Multi-Model Analysis of HTML Content}

\author*[1]{\fnm{Furkan} \sur{Çolhak}}\email{furkancolhak@stu.khas.edu.tr}
\author[1]{\fnm{Mert İlhan} \sur{Ecevit}}\email{mertilhan.ecevit@khas.edu.tr}
\equalcont{These authors contributed equally to this work.}
\author[1]{\fnm{Bilal Emir} \sur{Uçar}}\email{bilalucar@stu.khas.edu.tr}
\equalcont{These authors contributed equally to this work.}
\author[2,3]{\fnm{Reiner} \sur{Creutzburg}}\email{creutzburg@th-brandenburg.de}
\equalcont{These authors contributed equally to this work.}
\author[1]{\fnm{Hasan} \sur{Dağ}}\email{hasan.dag@khas.edu.tr}
\equalcont{These authors contributed equally to this work.}

\affil*[1]{\orgdiv{CCIP, Center for Cyber Security and Critical Infrastructure Protection}, \orgname{Kadir Has University}, \orgaddress{\postcode{34083}, \state{Istanbul}, \country{Turkey}}}

\affil[2]{\orgdiv{Berlin School of Technology}, \orgname{SRH Berlin University of Technology}, \orgaddress{ \postcode{D-10587}, \state{Berlin}, \country{Germany}}}

\affil[3]{\orgdiv{Department of Informatics and Media}, \orgname{Technische Hochschule Brandenburg}, \orgaddress{ \postcode{D-14770}, \state{Brandenburg}, \country{Germany}}}


\abstract{The way we communicate and work has changed significantly with the rise of the Internet. While it has opened up new opportunities, it has also brought about an increase in cyber threats. One common and serious threat is phishing, where cybercriminals employ deceptive methods to steal sensitive information. This study addresses the pressing phishing issue by introducing an advanced detection model that meticulously focuses on HTML content.

Our proposed approach integrates a specialized Multilayer Perceptron (MLP) model for structured tabular data and two pretrained Natural Language Processing (NLP) models for analyzing textual features such as page titles and content. The embeddings from these models are harmoniously combined through a novel fusion process. The resulting fused embeddings are then input into a linear classifier. Recognizing the scarcity of recent datasets for comprehensive phishing research, our contribution extends to creating an up-to-date dataset, which we openly share with the community. The dataset is meticulously curated to reflect real-life phishing conditions, ensuring relevance and applicability.

The research findings highlight the effectiveness of the proposed approach, with the CANINE demonstrating superior performance in analyzing page titles and the RoBERTa excelling in evaluating page content. The fusion of two NLP and one MLP model, termed MultiText-LP, achieves impressive results, yielding a 96.80 F1 score and a 97.18 accuracy score on our research dataset. Furthermore, our approach outperforms existing methods on the Aljofey's HTML dataset, showcasing its efficacies.}

\keywords{Phishing, Malicious Websites, Cybersecurity, Natural Language Processing, Multilayer Perceptron, Feature Extraction, Ensembles, Machine Learning in Cybersecurity, Pretrained Model}

\maketitle

\section{Introduction}\label{sec1}
The Internet Revolution began in the late 20th century and has significantly changed how we communicate, work, and live \cite{pencarelli2020digital}. It has connected people worldwide, providing unprecedented access to information, resources, and opportunities. However, along with the benefits, there is a growing challenge of internet dependence and an increase in cyber-attacks \cite{kamruzzaman2016plight}.

One notable threat in this digital age is phishing, a deceitful practice where cybercriminals trick individuals into revealing sensitive information like usernames, passwords, or financial details \cite{katkuri2018indian}. Phishing techniques have become sophisticated, utilizing deceptive emails, fake websites, and malicious software. Staying vigilant and informed is crucial as cybercriminals continuously evolve their strategies.

Phishing has become the most common form of cybercrime today. According to the Anti-Phishing Working Group (APWG), in the second quarter of 2023, 1,286,208 phishing attacks were observed—the third-highest quarterly total on record \cite{apwg2023}. The average wire transfer amount in Business Email Compromise (BEC) attacks surged to \$293,359, a 57\% increase from Q1. Additionally, an estimated 3.4 billion phishing emails are sent daily, with 1,350,037 reported attacks in Q4 2022 alone. Notably, 95\% of social engineering attack motivations are financially driven, emphasizing the high stakes involved \cite{apwg2023}.

Various software-based detection techniques are utilized in the ongoing fight against phishing attacks. Blacklists \cite{prakash2010phishnet} and whitelists \cite{jain2016novel} serve as standard tools, but their effectiveness may be limited in identifying new or subtly modified phishing sites. Machine learning-based approaches leverage heuristic features like URL \cite{rao2020catchphish} \cite{prabakaran2023enhanced},  HTML content \cite{jain2019machine} \cite{rao2022wordembedding} \cite{opara2020htmlphish}, website traffic, search engine data, and WHOIS records for classification, enhancing overall efficiency. Third-party services contribute valuable data to the detection process, including page rank, search engine indexing, and WHOIS \cite{cheng2022detecting} \cite{kuyama2016method}. Visual similarity-based heuristics compare new websites with pre-stored signatures based on screenshots, font styles, images, page layouts, and logos \cite{abdelnabi2020visualphishnet} \cite{chen2020intelligent}.

Even though there are many studies on spotting phishing using HTML, there are only a few recent datasets that anyone can use. Without Aljofey’s work \cite{aljofey2022effective}, it’s hard to find new HTML content datasets to compare how we’re doing. In our research, we collect and create data relevant to real-life phishing conditions. For detailed information about the dataset, please go to the Dataset section in our study (\ref{subsec3.1}). Additionally, we will share this dataset to contribute to the literature.

In our research, we’ve developed an advanced phishing detection model, focusing exclusively on HTML content. Our dataset includes both textual and tabular features. A specialized Multilayer Perceptron (MLP) model is employed for tabular data. In contrast, distinct pretrained Natural Language Processing (NLP) models, including transformers like BERT, RoBERTa, ALBERT, and CANINE, are utilized for processing textual features such as page titles and content. 

The embeddings from these models are brought together through a combination process. The combined embeddings are then input into a linear classifier, providing a comprehensive decision solely based on HTML content for effective phishing detection. This methodology combines tabular and textual feature processing, offering a potential aid in identifying phishing activities.
\vspace{2mm}

Our research has made contributions that can be outlined as:

\begin{itemize}[left=0.5in]
    \item Up-to-Date Dataset Creation and Open Sharing
    \item Innovative Fusion of pretrained Models and MLP for Phishing Detection
\end{itemize}

The remaining sections of this study are organized as follows: Section \ref{sec2} reviews related work. Section \ref{sec3} details the methodology employed, encompassing the construction of datasets, the extraction of features, and the development of models. Section \ref{sec4} is dedicated to presenting and analyzing the results, covering evaluation metrics and model performance. Section \ref{sec5} discusses limitations. Finally, Section \ref{sec6} concludes the study, summarizing key findings and insights.

\section{Related Work}\label{sec2}

This section overviews various phishing detection approaches categorized into URL-based, URL-HTML hybrid, pretrained-based, MLP-based, and combined MLP+NLP methods. It highlights vital studies and methodologies within each category, emphasizing the evolving landscape of anti-phishing techniques.

\textbf{URL-Based Approaches:}

Jain and Gupta (2016) introduced a client-side approach with an auto-updated white list, enhancing protection against phishing attacks. Their method provides individual users with an auto-updated white list, achieving an 86.02\% true positive rate with less than 1.48\% false negatives \cite{jain2016novel}. Prakash et al. (2010) presented PhishNet, utilizing predictive blacklisting for phishing attack detection. PhishNet employs heuristics to identify simple modifications of known phishing sites, resulting in the discovery of around 18,000 new phishing URLs with low false positive (3\%) and false negative (5\%) rates \cite{prakash2010phishnet}. Rao et al. (2020) developed CatchPhish, a lightweight application predicting URL legitimacy without website visits. The approach predicts URL legitimacy using features such as hostname, full URL, Term Frequency-Inverse Document Frequency (TF-IDF), and phish-hinted words, achieving high accuracy on benchmark datasets \cite{rao2020catchphish}. Prabakaran et al. (2023) proposed an enhanced mechanism using variational autoencoders for identifying malicious URLs.

Their approach integrates Variational Autoencoders (VAE) and deep neural networks (DNN) to automatically extract inherent features of raw URLs, achieving a maximum accuracy of 97.45\% \cite{prabakaran2023enhanced}.

\textbf{URL-HTML Hybrid Approaches:}
Jain and Gupta (2019) proposed a machine learning-based approach for phishing detection, analyzing hyperlinks within the HTML source code. The approach introduces 12 categories of hyperlink-specific features, achieving an accuracy surpassing 98.4\% on a logistic regression classifier \cite{jain2019machine}. Rao et al. (2022) applied word embedding and machine learning to detect phishing websites, focusing on domain-specific text. Their multimodal approach achieved a remarkable accuracy of 99.34\% using domain-specific text \cite{rao2022wordembedding}. Aljabri et al. (2022) assessed lexical, network, and content-based features for detecting malicious URLs. The study identifies Naïve Bayes (NB) as the most effective model, achieving an accuracy of 96\% \cite{aljabri2022assessment}. Opara et al. (2020) introduced HTMLPhish, a deep learning-based approach for phishing web page detection through HTML analysis. HTMLPhish, using Convolutional Neural Networks (CNNs), achieved high accuracy and a true positive rate of over 93\% \cite{opara2020htmlphish}.
Aljofey et al. (2022) presented a practical detection approach, combining URL and HTML features for phishing websites. Using character-level Term Frequency-Inverse Document Frequency (TF-IDF) features, their approach achieved high accuracy on both the authors’ dataset and a benchmark dataset \cite{aljofey2022effective}.

\textbf{Pretrained-based Approaches:}
Haynes et al. (2021) addressed phishing detection on mobile devices using NLP transformers, demonstrating lightweight URL-based phishing detection. Leveraging BERT and ELECTRA transformers, their approach showcases advantages in minimal training time, ease of updates, and real-time applicability on mobile devices \cite{haynes2021lightweight}. Wang et al. (2023) proposed a lightweight multi-view learning approach using transformers with a mixture of experts for effective phishing attack detection. The method incorporates URL analysis, attributes, content, and behavioral information, outperforming state-of-the-art approaches \cite{wang2023lightweight}. Misra and Rayz (2022) adapted pretrained language models to detect phishing emails, showcasing improvements over prior work. Their language models achieved nearly perfect performance on in-domain data and relative improvements on out-of-domain emails \cite{misra2022lms}.

\textbf{MLP-based Approaches:}
Odeh et al. (2020) developed an efficient prediction model using multilayer perceptron (MLP) to detect phishing websites. Their model aimed to efficiently predict phishing websites, contributing to improved internet security \cite{odeh2020efficient}. Belfedhal and Belfedhal (2022) presented a lightweight phishing detection system based on machine learning and URL features, achieving high accuracy using MLP. The multilayer perceptron (MLP) model emerged as the most effective, showcasing robust performance in automatically detecting phishing web pages \cite{belfedhal2022lightweight}.

\textbf{Combined MLP+NLP Approaches:}
In recent years, the combination of Multilayer Perceptron (MLP) and Natural Language Processing (NLP) has demonstrated its effectiveness across various domains. This synergy has been particularly evident in the works of Yin et al. (2020) and Ramos et al. (2022). Yin et al. introduced TaBERT, a pretrained language model designed for semantic parsing over both textual and tabular data, showcasing the versatility of MLP and NLP integration \cite{yin2020tabert}. Meanwhile, Ramos et al. leveraged BERT and numerical variables to classify injury leave based on accident descriptions, highlighting the potential of combining advanced NLP techniques with numerical features in a different domain \cite{ramos2022combining}.

However, despite the success of MLP and NLP integration in various applications, the utilization of this combined approach in cybersecurity is very limited. Düzgün et al. (2023) proposed a network intrusion detection system that uniquely combines tabular and text-based features. Integrating deep learning and pretrained transformer methodologies, their approach stands out for achieving superior accuracy in detecting network anomalies \cite{duzgun2023network}.
\vspace{1mm}

Examining existing literature underscores the unmet need for a specialized and comprehensive approach to phishing detection, explicitly focusing on HTML content. While various studies have advanced phishing detection in diverse aspects, there is a notable scarcity of exclusively HTML-based solutions. Our work pioneers a solution by intricately analyzing HTML structures. Our approach offers a distinctive strategy by seamlessly integrating embeddings from an MLP model and leveraging state-of-the-art pretrained NLP models. Successfully identifying phishing activities, our hybrid model substantially advances HTML-centric phishing detection by integrating two pretrained NLP models and MLP approaches.

\section{Methodology}\label{sec3}
\subsection{Dataset}\label{subsec3.1}
For the dataset, we compiled information from two distinct sources. Benign samples were sourced from the Alexa dataset, explicitly focusing on selecting initial URLs, notably the top 2000, and supplemented by randomly chosen benign URLs. Phishing URLs were gathered from OpenPhish. HTML content extraction from these URLs resulted in our dataset, comprising 36,848 instances of benign HTML content and 28,747 instances of phishing HTML content.

Furthermore, we incorporated the Aljofey's dataset, a widely used resource in existing literature. This dataset serves as a basis for comparing approaches, particularly in Aljofey’s work \cite{aljofey2022effective}, allowing us to assess the performance of our approach to others, with a specific emphasis on the analysis of HTML content.
Table 1 provides a detailed overview of the data distribution within our research and Aljofey's dataset. The research dataset is available on Google Drive as MTLP Dataset \cite{dataset_reference}.

\begin{table}[H]
  \centering
  \caption{Data Distribution in the MTLP and Aljofey's Datasets}
  \begin{tabular}{lccc}
    \toprule
    & \textbf{Benign Samples} & \textbf{Phishing Samples} & \textbf{Total} \\
    \midrule
    \textbf{MTLP Dataset \cite{dataset_reference}} & 36,848 & 28,747 & 65,595 \\
    \textbf{Aljofey's Dataset \cite{aljofey2022effective}} & 32,972 & 27,280 & 60,252 \\
    \bottomrule
  \end{tabular}
\end{table}

\subsection{Feature Extraction}\label{subsec3.2}
Our webpage analysis employs an extensive feature extraction process, encompassing key information from various aspects. In the realm of textual features (C1, C2), “page\_title” and “page\_content” offer insights into the webpage’s title and content, crucial for Natural Language Processing (NLP) models (NLP-1 and NLP-2). 
\vspace{1mm}

On the numeric side, features from C3 to C13 introduce diverse characteristics. For example, “hyperlink\_count” (C3) quantifies the total number of hyperlinks, “internal\_count” (C4) specifically tallies internal links, and “external\_count” (C5) enumerates external links. Additionally, C6 to C9 encapsulates counts of external CSS and JavaScript files, identification of foreign external CSS, and counts of obfuscated JavaScript files, contributing collectively to a unified numeric analysis.
\vspace{1mm}

The feature “suspicious\_form\_link” (C10) under login form features indicates links associated with a suspicious login form. Furthermore, C11 to C13 encompass aspects related to page structure and content, including “common\_page\_ratio,” “common\_page\_footer,” and “has\_meta\_description.” These numeric features collectively provide a comprehensive view, aligning with a Multilayer Perceptron (MLP) model for integrated and effective webpage assessment.
\vspace{1mm}

The extraction of these features draws inspiration from several studies \cite{aljofey2022effective} \cite{opara2020htmlphish} \cite{jain2019machine}. Table 2 illustrates the categories of features, their names, and the models to which they are related.

\begin{table}[H]
\footnotesize
\centering
\renewcommand{\arraystretch}{1.3}
\caption{Overview of Extracted Features}
\label{tab:webpage_features}
\begin{tabular}{|l|l|l|}
\hline
\textbf{Category} & \textbf{Feature Name} & \textbf{Related Model} \\ \hline
\multirow{2}{*}{Textual Feature 1 (C1)} & \multirow{2}{*}{page\_title} & \multirow{2}{*}{NLP-1} \\
& & \\ \hline
Textual Feature 2 (C2) & page\_content & NLP-2 \\ \hline
\multirow{3}{*}{Hyperlink Features (C3, C4, C5)} & \multirow{3}{*}{\begin{tabular}[c]{@{}l@{}}hyperlink\_count,\\ internal\_count,\\ external\_count\end{tabular}} & \multirow{3}{*}{MLP} \\
& & \\
& & \\ \hline
\multirow{4}{*}{CSS and JS Features (C6, C7, C8, C9)} & \multirow{4}{*}{\begin{tabular}[c]{@{}l@{}}external\_css\_count,\\ external\_js\_count,\\ foreign\_external\_css,\\ obfuscated\_js\_count\end{tabular}} & \multirow{4}{*}{MLP} \\
& & \\
& & \\
& & \\ \hline
Login Form Features (C10) & suspicious\_form\_link & MLP \\ \hline
\multirow{3}{*}{Page Structure and Content Features (C11, C12, C13)} & \multirow{3}{*}{\begin{tabular}[c]{@{}l@{}}common\_page\_ratio,\\ common\_page\_footer,\\ has\_meta\_description\end{tabular}} & \multirow{3}{*}{MLP} \\
& & \\
& & \\ \hline
\end{tabular}
\end{table}

\subsection{Model Development}\label{subsec3.3}

\subsubsection{Multilayer Perceptron (MLP)}\label{subsec3.3.1}

An MLP is an artificial neural network consisting of multiple layers of nodes \cite{taud2018mlp}, also known as neurons or perceptrons. It comprises an input layer, one or more hidden layers, and an output layer. Neurons in one layer are connected to neurons in the next layer by weights, and each connection may have an associated bias \cite{aitkin2003statistical}. The activation function introduces non-linearity to the model, allowing it to learn complex patterns in the data.

\begin{table}[H]
  \centering
  \caption{MLP Structure}
  \label{tab:mlp-architecture}
  \small
  \begin{tabular}{|c|c|c|c|}
    \toprule
    \textbf{Layer} & \textbf{Dimensions} & \textbf{Activation} & \textbf{Reg.} \\
    \midrule
    Input & $[input\_dim]$ & & \\
    FC1 & $[input\_dim, 1024]$ & LeakyReLU (0.1) & BN \\
    Dropout & $p=0.2$ & & \\
    FC2 & $[1024, 2056]$ & LeakyReLU (0.1) & BN \\
    Dropout & $p=0.2$ & & \\
    FC3 & $[2056, 512]$ & LeakyReLU (0.1) & BN \\
    Dropout & $p=0.2$ & & \\
    FC4 & $[512, 16]$ & LeakyReLU (0.1) & BN \\
    Dropout & $p=0.3$ & & \\
    FC5 & $[16, output\_dim]$ & & \\
    \bottomrule
  \end{tabular}
\end{table}
\textbf{Note:} $[input\_dim]=11$ corresponds to the number of numeric columns in the input data, and $[output\_dim]=2$ represents benign/phishing classification.

\vspace{1mm}
Table \ref{tab:mlp-architecture} represents the structure of the MLP utilized in this research. The model incorporates multiple fully connected layers with LeakyReLU activation and Batch Normalization for improved training stability. Dropout layers are strategically placed to mitigate overfitting. The dimensions of the hidden layers progressively decrease, facilitating the extraction of hierarchical features. The final layer outputs predictions with dimensions corresponding to the specified $output_dim$. This architecture is designed to capture complex relationships within the input data and generalize well to achieve the research objectives.

\subsubsection{Pretrained NLP Models}\label{subssec3.3.2}
Transformer models, introduced by Vaswani et al. in 2017 \cite{vaswani2017attention}, have significantly advanced neural network architecture, particularly in Natural Language Processing (NLP). Their innovative use of self-attention mechanisms enables dynamic weighing of input elements, capturing long-range dependencies and relationships in sequential data like text or speech. These transformers are widely adopted across NLP applications, excelling in language translation, text classification (sentiment analysis, topic classification, document categorization), and named entity recognition.

Pretrained transformer models, a cornerstone in NLP, come pretrained on extensive datasets, offering general language representations. They serve as a foundation for fine-tuning specific tasks. They provide benefits such as capturing intricate linguistic patterns as starting points for new models and facilitating efficient training by leveraging pretrained weights.

ALBERT (A Lite BERT) stands out as an efficient variant in the realm of pretrained transformer models \cite{lan2019albert}. It addresses efficiency concerns without compromising performance through techniques like factorized embedding parameterization and cross-layer parameter sharing. BERT, a pioneering model, introduced bidirectional context to NLP, enhancing word relationships and semantics comprehension \cite{devlin2018bert}. CANINE (Character Architecture with No Tokenization in Neural Encoders) is another pretrained encoder model designed to overcome the limitations of traditional tokenization techniques. Unlike conventional models, CANINE uses neural encoders that encode the sequence of characters or sub-words without explicitly tokenizing the input data \cite{clark2021canine}.

RoBERTa \cite{tan2022roberta-lstm}  (Robustly Optimized BERT Approach with Pre-training) refines BERT’s successes by modifying training strategies and eliminating the sentence prediction objective, achieving state-of-the-art performance across diverse NLP benchmarks. In our research, we utilize these pretrained models to compare their effectiveness in handling and extracting information from different textual features within the dataset, specifically the page\_title and page\_content columns.

\subsubsection{MultiText-LP}\label{subsec3.3.3}
Introducing the MultiText-LP model, a cutting-edge fusion of two NLP pretrained models and one MLP model designed to significantly boost classification performance across a wide range of data formats.

Tailored specifically for handling numeric and categorical data, the MultiText-LP model incorporates an MLP to effectively uncover intricate relationships and patterns within the dataset. At the same time, it integrates two distinct NLP pretrained models for processing text information. NLP-1 is dedicated to features like ‘page\_title,’ while NLP-2 handles’ page\_content,’ as illustrated in Figure 1.

\begin{figure}[H]
    \centering
    \includegraphics[width=1\linewidth]{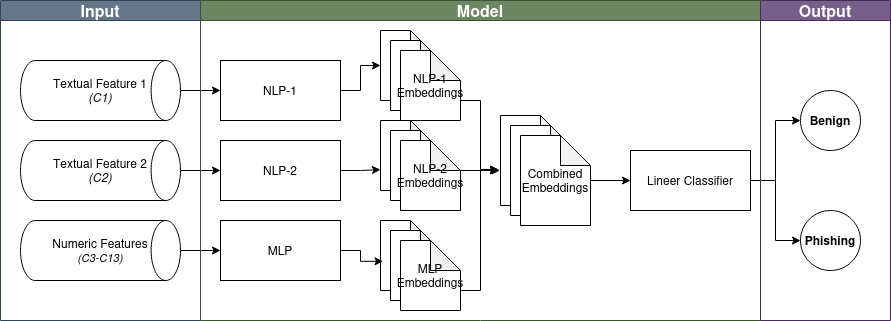}
    \caption{MultiText-LP Model Structure}
    \label{fig:enter-label}
\end{figure}

The synergy of these diverse models within the MultiText-LP architecture brings unparalleled versatility and performance, surpassing the capabilities of individual models. This combination allows the MultiText-LP model to adeptly handle numeric, categorical, and text data, making it a powerful tool for addressing complex classification challenges.

In the MultiText-LP workflow, embeddings from different models harmonize to create a comprehensive representation, seamlessly feeding into a linear classifier. The result of this innovative approach indicates whether the input data is benign or potentially phishing. The MultiText-LP model showcases its adaptability and potency in tackling complex classification tasks with remarkable efficacy.

\section{Results}\label{sec4}
\subsection{Evaluation Metrics}\label{subsec4.1}

To evaluate our approach, we employed various metrics, including accuracy, precision, recall, and F1-score. A comprehensive overview of these metrics and detailed descriptions are provided in Table \ref{tab:phishing_metrics}.
In assessing our model, we designate the F1-score as the primary indicator due to its relevance in determining phishing websites.
\vspace{1mm}

The F1-score, the harmonic mean of precision and recall, ensures a balanced evaluation. This is critical in phishing detection, where minimizing false positives (precision) and capturing a significant portion of actual phishing instances (recall) is imperative. This metric is especially pertinent in the high-stakes nature of phishing identification, where the consequences of misclassifying benign content or overlooking genuine threats are substantial.
\vspace{1mm}

Given its importance in measuring overall performance, we consider accuracy a secondary metric. While accuracy provides a general assessment of correct predictions, the F1-score takes precedence in our evaluation, offering a more nuanced and targeted measure of our model’s efficacy in the specific challenge of phishing detection.

\begin{table}[H]
\centering
\caption{Metrics Formulas and Descriptions}
\label{tab:phishing_metrics}
\begin{tabular}{|c|c|m{0.5\textwidth}|}
\hline
\textbf{Metric} & \textbf{Formula} & \textbf{Description} \\
\hline
\textit{Accuracy} & $\frac{TP + TN}{TP + TN + FP + FN} \times 100$ & Percentage of correctly classified phishing and benign websites out of the total instances. \\
\hline
\textit{Precision} & $\frac{TP}{TP + FP}$ & Proportion of correctly predicted phishing websites among those predicted as phishing. \\
\hline
\textit{Recall} & $\frac{TP}{TP + FN}$ & Proportion of correctly predicted phishing websites out of all actual phishing websites. \\
\hline
\textit{F1-Score} & $2 \times \frac{\textit{Precision} \times \textit{Recall}}{\textit{Precision} + \textit{Recall}}$ & Harmonic mean of precision and recall, providing a balanced measure that considers both false positives and false negatives. \\
\hline
\end{tabular}
\end{table}

\subsection{Experimental Setup}\label{subsec4.2}

Experiments were conducted on a system powered by an Intel Core i9-10900x processor, featuring a base clock speed of 3.70 GHz. Computational tasks were enhanced with an NVIDIA RTX A4000 GPU, equipped with 16GB of high-speed video memory featuring an error-correction code (ECC). 
\vspace{1mm}

On the other hand, the research’s dataset was randomly partitioned into 70\% training, 15\% validation, and 15\% testing sets. In contrast, for a fair comparison, the Aljofey's dataset was divided into 80\% training and 20\% testing, following Aljofey’s approach \cite{aljofey2022effective}.

\subsection{Model Performance}\label{subsec4.3}

In this section, we present a comprehensive analysis of the performance of our proposed MultiText-LP model across various dimensions. We conduct a meticulous examination by comparing our model against individual pretrained NLP models, a standalone MLP model, and an existing approach, Aljofey’s \cite{aljofey2022effective}, on the research dataset. Our goal is to showcase the enhanced classification capabilities achieved through the fusion of NLP and MLP models.

\begin{table}[H]
  \centering
  \caption{Performance Comparison of pretrained Models}
  \label{tab:nlpmodel-comparison}
  \begin{tabular}{llcc}
    \toprule
    \textbf{Feature} & \textbf{Pretrained Model} & \textbf{F1-Score (\%)} & \textbf{Accuracy (\%)} \\
    \midrule
    \multirow{6}{*}{\textbf{NLP-1 (Page Title)}} & ALBERT & 92.85 & 93.43 \\
                                    & BERT & 93.14 & 93.72 \\
                                    & \cellcolor[HTML]{9AFF99}\textbf{CANINE} & \cellcolor[HTML]{9AFF99}\textbf{93.33} & \cellcolor[HTML]{9AFF99}\textbf{93.84} \\
                                    & RoBERTa & 92.92 & 93.56 \\
    \midrule
    \multirow{6}{*}{\textbf{NLP-2 (Page Content)}} & ALBERT & 94.59 & 95.08 \\
                                    & BERT & 94.79 & 95.26 \\
                                    & CANINE & 96.29 & 96.69 \\
                                    & \cellcolor[HTML]{9AFF99}\textbf{RoBERTa} & \cellcolor[HTML]{9AFF99}\textbf{96.38} & \cellcolor[HTML]{9AFF99}\textbf{96.76}
  \end{tabular}
\end{table}

\vspace{2mm}
To contextualize the performance of MultiText-LP, we begin by summarizing the performance of various pretrained models on specific features (Table \ref{tab:nlpmodel-comparison}). For ‘NLP-1 (Page Title),’ the CANINE model outperforms others, boasting an F1-Score of 93.33\% and an accuracy of 93.84\%. In the ‘NLP-2 (Page Content)’ category, RoBERTa demonstrates superior performance with an F1-Score of 96.38\% and an accuracy of 96.76\%.

\begin{table}[H]
  \centering
  \caption{Comparison between models of Accuracy and F1-Scores}
  \label{tab:model-comparison}
  \large  
  \begin{tabular}{lcc}
    \toprule
    \textbf{Model} & \textbf{F1-Score (\%)} & \textbf{Accuracy (\%)} \\
    \midrule
    MLP & 88.83 & 89.92 \\
    NLP-1 (CANINE) & 93.33 & 93.84 \\
    NLP-2 (RoBERTa) & 96.38 & 96.76 \\
    \cellcolor[HTML]{9AFF99}\textbf{MultiText-LP} & \cellcolor[HTML]{9AFF99}\textbf{96.80} & \cellcolor[HTML]{9AFF99}\textbf{97.18} \\
    \bottomrule
  \end{tabular}
\end{table}

\vspace{2mm}
Table \ref{tab:model-comparison} provides a holistic comparison, where the standalone MLP model achieves an F1-Score of 88.83\% and an accuracy of 89.92\%. The NLP-1 and NLP-2 models individually surpass the MLP’s performance, showcasing F1-Scores of 93.33\%/96.38\% and accuracies of 93.84\%/96.76\%, respectively. The MultiText-LP model, a fusion of NLP-1, NLP-2, and MLP models, outshines them all with an F1-Score of 96.80\% and an accuracy of 97.18\%, illustrating the synergistic effect of combining both approaches.

\definecolor{positive}{RGB}{154, 255, 153}
\definecolor{negative}{RGB}{255, 192, 192}

\begin{table}[H]
  \centering
  \caption{Approach Performance Comparison on the \textit{Aljofey's} Dataset}
  \resizebox{\linewidth}{!}{%
  \begin{tabular}{lcccc}
    \toprule
    \textbf{Model} & \textbf{Accuracy (\%)} & \textbf{Precision (\%)} & \textbf{Recall (\%)} & \textbf{F1-Score (\%)} \\
    \midrule
    Aljofey's Approach \cite{aljofey2022effective} & 89.00 & 88.20 & 87.68 & 87.94 \\
    \textbf{MultiText-LP} & \textbf{89.58}  & \textbf{89.61} & \textbf{87.03} & \textbf{88.17} \\
    \midrule
    \textbf{Difference} & \cellcolor{positive}\textbf{+0.58} & \cellcolor{positive}\textbf{+1.41} & \cellcolor{negative}\textbf{-0.65} & \cellcolor{positive}\textbf{+0.23} \\
    \bottomrule
  \end{tabular}%
  }
  \label{tab:approach_comparison}
\end{table}

A detailed evaluation in Table \ref{tab:approach_comparison} focuses on HTML content within the challenging Aljofey's dataset. Aljofey’s Approach \cite{aljofey2022effective} demonstrates commendable metrics, yet the MultiText-LP model showcases superior results with an accuracy of 89.58\%, precision of 89.61\%, and F1-Score of 88.17\%. Despite a slight adjustment in recall (87.03\%), our analysis reveals positive increments in accuracy (+0.58\%), precision (+1.41\%), and F1-Score (+0.23\%). This emphasizes the enhanced efficacy of the MultiText-LP model in HTML content classification compared to Aljofey’s Approach.
\vspace{2mm}

Our innovative MultiText-LP model emerges as a powerful tool for HTML content classification, exhibiting superior performance compared to individual NLP models, a standalone MLP model, and a state-of-the-art approach. The fusion of two NLP and MLP demonstrates a synergistic effect, enhancing classification capabilities. With promising results on the challenging Aljofey's dataset, our approach holds significant potential for applications in the broader HTML content analysis and classification field.

\section{Limitations}\label{sec5}
While our approach demonstrates success in its objectives, it is not without limitations. Using two pretrained and one MLP model simultaneously requires a powerful GPU like A4000. Standard GPUs may struggle with their size, making training and deployment less efficient, especially in places with limited resources.

Another limitation is the limited availability of datasets, which includes HTML content. We can only access the Aljofey's dataset from Aljofey’s study \cite{aljofey2022effective}. This scarcity makes it challenging to find comparable datasets and hinders direct comparisons with other approaches. Moreover, there is a lack of prior work focusing explicitly on dealing with HTML content. This absence of existing methodologies for working solely with HTML further complicates our efforts to benchmark and evaluate our approach compared to established methods.

\section{Conclusion and Future Work}\label{sec6}
As phishing websites get more sophisticated daily, detecting them has become a significant challenge. Our paper focuses on detecting phishing websites by looking at their HTML content.

While research efforts have focused on detecting phishing websites, a notable gap exists in analyzing HTML content. Several studies have explored URLs and HTML, but a noteworthy gap exists in research focusing on analyzing HTML content alone. Because of that, dataset accessibility is minimal. To address this gap, we curated an up-to-date dataset that reflects real-life phishing conditions, ensuring its relevance and applicability. This dataset, shared openly with the community, contributes to the existing literature and provides a valuable resource for comprehensive phishing research.

We proposed the MultiText-LP model, a novel fusion of two pretrained Natural Language Processing (NLP) models (CANINE and RoBERTa) and a specialized Multilayer Perceptron (MLP) model. Results illustrate the superior performance of the MultiText-LP model compared to individual pretrained NLP models and a standalone MLP model. With an impressive F1 score of 96.80\% and an accuracy of 97.18\ on our research dataset, the MultiText-LP model demonstrates its efficacy in accurately identifying phishing activities. 

Our innovative MultiText-LP model emerges as a powerful tool for HTML content classification, exhibiting superior performance compared to individual NLP models, a standalone MLP model, and a state-of-the-art approach. The fusion of two NLP and MLP demonstrates a synergistic effect, enhancing classification capabilities. With promising results on the challenging Aljofey's dataset, our approach holds significant potential for applications in the broader HTML content analysis and classification field.

We will integrate URL, HTML content, and WHOIS data in our upcoming projects, leveraging this combined approach. However, this integration may pose computational challenges. We intend to optimize the model’s efficiency and reduce its overall capacity to address this.

\section{Acknowledgments}
This work was supported partially by the European Union in the framework of ERASMUS MUNDUS, Project CyberMACS (Project \#101082683) (\url{https://cybermacs.eu}).\\

\bibliography{sn-bibliography}


\begin{thebibliography}{33}
\ifx \bisbn   \undefined \def \bisbn  #1{ISBN #1}\fi
\ifx \binits  \undefined \def \binits#1{#1}\fi
\ifx \bauthor  \undefined \def \bauthor#1{#1}\fi
\ifx \batitle  \undefined \def \batitle#1{#1}\fi
\ifx \bjtitle  \undefined \def \bjtitle#1{#1}\fi
\ifx \bvolume  \undefined \def \bvolume#1{\textbf{#1}}\fi
\ifx \byear  \undefined \def \byear#1{#1}\fi
\ifx \bissue  \undefined \def \bissue#1{#1}\fi
\ifx \bfpage  \undefined \def \bfpage#1{#1}\fi
\ifx \blpage  \undefined \def \blpage #1{#1}\fi
\ifx \burl  \undefined \def \burl#1{\textsf{#1}}\fi
\ifx \doiurl  \undefined \def \doiurl#1{\url{https://doi.org/#1}}\fi
\ifx \betal  \undefined \def \betal{\textit{et al.}}\fi
\ifx \binstitute  \undefined \def \binstitute#1{#1}\fi
\ifx \binstitutionaled  \undefined \def \binstitutionaled#1{#1}\fi
\ifx \bctitle  \undefined \def \bctitle#1{#1}\fi
\ifx \beditor  \undefined \def \beditor#1{#1}\fi
\ifx \bpublisher  \undefined \def \bpublisher#1{#1}\fi
\ifx \bbtitle  \undefined \def \bbtitle#1{#1}\fi
\ifx \bedition  \undefined \def \bedition#1{#1}\fi
\ifx \bseriesno  \undefined \def \bseriesno#1{#1}\fi
\ifx \blocation  \undefined \def \blocation#1{#1}\fi
\ifx \bsertitle  \undefined \def \bsertitle#1{#1}\fi
\ifx \bsnm \undefined \def \bsnm#1{#1}\fi
\ifx \bsuffix \undefined \def \bsuffix#1{#1}\fi
\ifx \bparticle \undefined \def \bparticle#1{#1}\fi
\ifx \barticle \undefined \def \barticle#1{#1}\fi
\bibcommenthead
\ifx \bconfdate \undefined \def \bconfdate #1{#1}\fi
\ifx \botherref \undefined \def \botherref #1{#1}\fi
\ifx \url \undefined \def \url#1{\textsf{#1}}\fi
\ifx \bchapter \undefined \def \bchapter#1{#1}\fi
\ifx \bbook \undefined \def \bbook#1{#1}\fi
\ifx \bcomment \undefined \def \bcomment#1{#1}\fi
\ifx \oauthor \undefined \def \oauthor#1{#1}\fi
\ifx \citeauthoryear \undefined \def \citeauthoryear#1{#1}\fi
\ifx \endbibitem  \undefined \def \endbibitem {}\fi
\ifx \bconflocation  \undefined \def \bconflocation#1{#1}\fi
\ifx \arxivurl  \undefined \def \arxivurl#1{\textsf{#1}}\fi
\csname PreBibitemsHook\endcsname

\bibitem[\protect\citeauthoryear{Pencarelli}{2020}]{pencarelli2020digital}
\begin{barticle}
\bauthor{\bsnm{Pencarelli}, \binits{T.}}:
\batitle{The digital revolution in the travel and tourism industry}.
\bjtitle{Information Technology \& Tourism}
\bvolume{22}(\bissue{3}),
\bfpage{455}--\blpage{476}
(\byear{2020}).
\bcomment{\url{https://doi.org/10.1007/s40558-019-00160-3}}
\end{barticle}
\endbibitem

\bibitem[\protect\citeauthoryear{Kamruzzaman et~al.}{2016}]{kamruzzaman2016plight}
\begin{barticle}
\bauthor{\bsnm{Kamruzzaman}, \binits{M.}},
\bauthor{\bsnm{Islam}, \binits{M.A.}},
\bauthor{\bsnm{Islam}, \binits{M.S.}},
\bauthor{\bsnm{Hossain}, \binits{M.S.}},
\bauthor{\bsnm{Hakim}, \binits{M.A.}}:
\batitle{Plight of youth perception on cybercrime in south asia}.
\bjtitle{American Journal of Information Science and Computer Engineering}
\bvolume{2}(\bissue{4}),
\bfpage{22}--\blpage{28}
(\byear{2016})
\end{barticle}
\endbibitem

\bibitem[\protect\citeauthoryear{Katkuri}{2018}]{katkuri2018indian}
\begin{barticle}
\bauthor{\bsnm{Katkuri}, \binits{S.}}:
\batitle{Indian cyber law}.
\bjtitle{International Journal of Advanced Research and Development}
\bvolume{3}(\bissue{1}),
\bfpage{640}--\blpage{644}
(\byear{2018})
\end{barticle}
\endbibitem

\bibitem[\protect\citeauthoryear{{Anti-Phishing Working Group (APWG)}}{2023}]{apwg2023}
\begin{botherref}
\oauthor{\bsnm{{Anti-Phishing Working Group (APWG)}}}:
Phishing Trends and Statistics: Q2 2023.
pp. 4-6.
\url{https://apwg.org/trendsreports/}
\end{botherref}
\endbibitem

\bibitem[\protect\citeauthoryear{Prakash et~al.}{2010}]{prakash2010phishnet}
\begin{bchapter}
\bauthor{\bsnm{Prakash}, \binits{P.}},
\bauthor{\bsnm{Kumar}, \binits{M.}},
\bauthor{\bsnm{Kompella}, \binits{R.R.}},
\bauthor{\bsnm{Gupta}, \binits{M.}}:
\bctitle{Phishnet: Predictive blacklisting to detect phishing attacks}.
In: \bbtitle{2010 Proceedings IEEE INFOCOM},
pp. \bfpage{1}--\blpage{5}
(\byear{2010}).
\doiurl{10.1109/INFCOM.2010.5462216}
\end{bchapter}
\endbibitem

\bibitem[\protect\citeauthoryear{Jain and Gupta}{2016}]{jain2016novel}
\begin{botherref}
\oauthor{\bsnm{Jain}, \binits{A.K.}},
\oauthor{\bsnm{Gupta}, \binits{B.B.}}:
A novel approach to protect against phishing attacks at client side using auto-updated white-list.
EURASIP Journal on Information Security,
1--11
(2016).
\url{https://doi.org/10.1186/s13635-016-0034-3}
\end{botherref}
\endbibitem

\bibitem[\protect\citeauthoryear{Rao et~al.}{2020}]{rao2020catchphish}
\begin{barticle}
\bauthor{\bsnm{Rao}, \binits{R.S.}},
\bauthor{\bsnm{Vaishnavi}, \binits{T.}},
\bauthor{\bsnm{Pais}, \binits{A.R.}}:
\batitle{Catchphish: detection of phishing websites by inspecting urls}.
\bjtitle{Journal of Ambient Intelligence and Humanized Computing}
\bvolume{11},
\bfpage{813}--\blpage{825}
(\byear{2020}).
\bcomment{\url{https://doi.org/10.1007/s12652-019-01311-4}}
\end{barticle}
\endbibitem

\bibitem[\protect\citeauthoryear{Prabakaran et~al.}{2023}]{prabakaran2023enhanced}
\begin{barticle}
\bauthor{\bsnm{Prabakaran}, \binits{M.K.}},
\bauthor{\bsnm{Sundaram}, \binits{P.M.}},
\bauthor{\bsnm{Chandrasekar}, \binits{A.D.}}:
\batitle{An enhanced deep learning‐based phishing detection mechanism to effectively identify malicious urls using variational autoencoders}.
\bjtitle{IET Information Security}
\bvolume{17}(\bissue{3}),
\bfpage{423}--\blpage{440}
(\byear{2023}).
\bcomment{\url{ https://doi.org/10.1049/ise2.12106}}
\end{barticle}
\endbibitem

\bibitem[\protect\citeauthoryear{Jain and Gupta}{2019}]{jain2019machine}
\begin{barticle}
\bauthor{\bsnm{Jain}, \binits{A.K.}},
\bauthor{\bsnm{Gupta}, \binits{B.B.}}:
\batitle{A machine learning based approach for phishing detection using hyperlinks information}.
\bjtitle{Journal of Ambient Intelligence and Humanized Computing}
\bvolume{10},
\bfpage{2015}--\blpage{2028}
(\byear{2019}).
\bcomment{\url{https://doi.org/10.1007/s12652-018-0798-z}}
\end{barticle}
\endbibitem

\bibitem[\protect\citeauthoryear{Rao et~al.}{2022}]{rao2022wordembedding}
\begin{botherref}
\oauthor{\bsnm{Rao}, \binits{R.S.}},
\oauthor{\bsnm{Umarekar}, \binits{A.}},
\oauthor{\bsnm{Pais}, \binits{A.R.}}:
Application of word embedding and machine learning in detecting phishing websites.
Telecommunication Systems,
1--13
(2022).
\url{https://doi.org/10.1007/s11235-021-00850-6}
\end{botherref}
\endbibitem

\bibitem[\protect\citeauthoryear{Opara et~al.}{2020}]{opara2020htmlphish}
\begin{bchapter}
\bauthor{\bsnm{Opara}, \binits{C.}},
\bauthor{\bsnm{Wei}, \binits{B.}},
\bauthor{\bsnm{Chen}, \binits{Y.}}:
\bctitle{Htmlphish: Enabling phishing web page detection by applying deep learning techniques on html analysis}.
In: \bbtitle{2020 International Joint Conference on Neural Networks (IJCNN)},
pp. \bfpage{1}--\blpage{8}
(\byear{2020}).
\doiurl{10.1109/IJCNN48605.2020.9207707}
\end{bchapter}
\endbibitem

\bibitem[\protect\citeauthoryear{Cheng et~al.}{2022}]{cheng2022detecting}
\begin{barticle}
\bauthor{\bsnm{Cheng}, \binits{Y.}}, \betal:
\batitle{Detecting malicious domain names with abnormal whois records using feature-based rules}.
\bjtitle{The Computer Journal}
\bvolume{65}(\bissue{9}),
\bfpage{2262}--\blpage{2275}
(\byear{2022}).
\bcomment{\url{https://doi.org/10.1093/comjnl/bxab062}}
\end{barticle}
\endbibitem

\bibitem[\protect\citeauthoryear{Kuyama et~al.}{2016}]{kuyama2016method}
\begin{bchapter}
\bauthor{\bsnm{Kuyama}, \binits{M.}},
\bauthor{\bsnm{Kakizaki}, \binits{Y.}},
\bauthor{\bsnm{Sasaki}, \binits{R.}}:
\bctitle{Method for detecting a malicious domain by using whois and dns features}.
In: \bbtitle{The Third International Conference on Digital Security and Forensics (DigitalSec2016)},
vol. \bseriesno{74}
(\byear{2016})
\end{bchapter}
\endbibitem

\bibitem[\protect\citeauthoryear{Abdelnabi et~al.}{2020}]{abdelnabi2020visualphishnet}
\begin{bchapter}
\bauthor{\bsnm{Abdelnabi}, \binits{S.}},
\bauthor{\bsnm{Krombholz}, \binits{K.}},
\bauthor{\bsnm{Fritz}, \binits{M.}}:
\bctitle{Visualphishnet: Zero-day phishing website detection by visual similarity}.
In: \bbtitle{Proceedings of the 2020 ACM SIGSAC Conference on Computer and Communications Security}
(\byear{2020}).
\bcomment{\url{https://doi.org/10.1145/3372297.3417233}}
\end{bchapter}
\endbibitem

\bibitem[\protect\citeauthoryear{Chen et~al.}{2020}]{chen2020intelligent}
\begin{barticle}
\bauthor{\bsnm{Chen}, \binits{J.-L.}},
\bauthor{\bsnm{Ma}, \binits{Y.-W.}},
\bauthor{\bsnm{Huang}, \binits{K.-L.}}:
\batitle{Intelligent visual similarity-based phishing websites detection}.
\bjtitle{Symmetry}
\bvolume{12}(\bissue{10}),
\bfpage{1681}
(\byear{2020}).
\bcomment{\url{https://doi.org/10.3390/sym12101681}}
\end{barticle}
\endbibitem

\bibitem[\protect\citeauthoryear{Aljofey}{2022}]{aljofey2022effective}
\begin{barticle}
\bauthor{\bsnm{Aljofey}, \binits{A.e.a.}}:
\batitle{An effective detection approach for phishing websites using url and html features}.
\bjtitle{Scientific Reports}
\bvolume{12}(\bissue{1}),
\bfpage{8842}
(\byear{2022}).
\bcomment{\url{https://doi.org/10.1038/s41598-022-10841-5}}
\end{barticle}
\endbibitem

\bibitem[\protect\citeauthoryear{Aljabri}{2022}]{aljabri2022assessment}
\begin{botherref}
\oauthor{\bsnm{Aljabri}, \binits{M.e.a.}}:
An assessment of lexical, network, and content-based features for detecting malicious urls using machine learning and deep learning models.
Computational Intelligence and Neuroscience
(2022).
\url{https://doi.org/10.1155/2022/3241216}
\end{botherref}
\endbibitem

\bibitem[\protect\citeauthoryear{Haynes et~al.}{2021}]{haynes2021lightweight}
\begin{barticle}
\bauthor{\bsnm{Haynes}, \binits{K.}},
\bauthor{\bsnm{Shirazi}, \binits{H.}},
\bauthor{\bsnm{Ray}, \binits{I.}}:
\batitle{Lightweight url-based phishing detection using natural language processing transformers for mobile devices}.
\bjtitle{Procedia Computer Science}
\bvolume{191},
\bfpage{127}--\blpage{134}
(\byear{2021}).
\bcomment{\url{https://doi.org/10.1016/j.procs.2021.07.040}}
\end{barticle}
\endbibitem

\bibitem[\protect\citeauthoryear{Wang}{2023}]{wang2023lightweight}
\begin{barticle}
\bauthor{\bsnm{Wang}, \binits{Y.e.a.}}:
\batitle{A lightweight multi-view learning approach for phishing attack detection using transformer with mixture of experts}.
\bjtitle{Applied Sciences}
\bvolume{13}(\bissue{13}),
\bfpage{7429}
(\byear{2023}).
\bcomment{\url{https://doi.org/10.3390/app13137429}}
\end{barticle}
\endbibitem

\bibitem[\protect\citeauthoryear{Misra and Rayz}{2022}]{misra2022lms}
\begin{bchapter}
\bauthor{\bsnm{Misra}, \binits{K.}},
\bauthor{\bsnm{Rayz}, \binits{J.T.}}:
\bctitle{Lms go phishing: Adapting pre-trained language models to detect phishing emails}.
In: \bbtitle{2022 IEEE/WIC/ACM International Joint Conference on Web Intelligence and Intelligent Agent Technology (WI-IAT)},
pp. \bfpage{135}--\blpage{142}
(\byear{2022}).
\doiurl{10.1109/WI-IAT55865.2022.00028}
\end{bchapter}
\endbibitem

\bibitem[\protect\citeauthoryear{Odeh}{2020}]{odeh2020efficient}
\begin{barticle}
\bauthor{\bsnm{Odeh}, \binits{A.M.M.A.R.e.a.}}:
\batitle{Efficient prediction of phishing websites using multilayer perceptron (mlp)}.
\bjtitle{Journal of Theoretical and Applied Information Technology}
\bvolume{98}(\bissue{16}),
\bfpage{3353}--\blpage{3363}
(\byear{2020})
\end{barticle}
\endbibitem

\bibitem[\protect\citeauthoryear{Belfedhal and Belfedhal}{2022}]{belfedhal2022lightweight}
\begin{bchapter}
\bauthor{\bsnm{Belfedhal}, \binits{A.E.}},
\bauthor{\bsnm{Belfedhal}, \binits{M.A.}}:
\bctitle{A lightweight phishing detection system based on machine learning and url features}.
In: \bbtitle{International Conference on Managing Business Through Web Analytics}
(\byear{2022}).
\bcomment{Springer. \url{https://doi.org/10.1007/978-3-031-06971-0_22}}
\end{bchapter}
\endbibitem

\bibitem[\protect\citeauthoryear{Yin et~al.}{2020}]{yin2020tabert}
\begin{botherref}
\oauthor{\bsnm{Yin}, \binits{P.}},
\oauthor{\bsnm{Zhang}, \binits{G.N.L.W.}},
\oauthor{\bsnm{Alex}}:
Tabert: Pretraining for joint understanding of textual and tabular data.
arXiv preprint arXiv:2005.08314
(2020).
\url{https://doi.org/10.48550/arXiv.2005.08314}
\end{botherref}
\endbibitem

\bibitem[\protect\citeauthoryear{Ramos et~al.}{2022}]{ramos2022combining}
\begin{botherref}
\oauthor{\bsnm{Ramos}, \binits{P.M.}}, et al.:
Combining bert with numerical variables to classify injury leave based on accident description.
Proceedings of the Institution of Mechanical Engineers, Part O: Journal of Risk and Reliability
(2022).
\url{https://doi.org/10.1177/1748006X221140194}
\end{botherref}
\endbibitem

\bibitem[\protect\citeauthoryear{Düzgün}{2023}]{duzgun2023network}
\begin{botherref}
\oauthor{\bsnm{Düzgün}, \binits{B.e.a.}}:
Network intrusion detection system by learning jointly from tabular and text‐based features.
Expert Systems,
13518
(2023)
\doiurl{10.1111/exsy.13518}
\end{botherref}
\endbibitem

\bibitem[\protect\citeauthoryear{{Çolhak, Furkan}}{2024}]{dataset_reference}
\begin{botherref}
\oauthor{\bsnm{{Çolhak, Furkan}}}:
MTLP Dataset.
\url{https://drive.google.com/file/d/1Lp3ueOd7AxmAl2Y0jJ2U2XlEFa6q8AcT/view?usp=drive_link}
\end{botherref}
\endbibitem

\bibitem[\protect\citeauthoryear{Taud and Mas}{2018}]{taud2018mlp}
\begin{bchapter}
\bauthor{\bsnm{Taud}, \binits{H.}},
\bauthor{\bsnm{Mas}, \binits{J.F.}}:
\bctitle{Multilayer perceptron (mlp)}.
In: \bbtitle{Geomatic Approaches for Modeling Land Change Scenarios},
pp. \bfpage{451}--\blpage{455}
(\byear{2018}).
\bcomment{\url{https://doi.org/10.1007/978-3-319-60801-3_27}}
\end{bchapter}
\endbibitem

\bibitem[\protect\citeauthoryear{Aitkin and Foxall}{2003}]{aitkin2003statistical}
\begin{barticle}
\bauthor{\bsnm{Aitkin}, \binits{M.}},
\bauthor{\bsnm{Foxall}, \binits{R.}}:
\batitle{Statistical modelling of artificial neural networks using the multi-layer perceptron}.
\bjtitle{Statistics and Computing}
\bvolume{13},
\bfpage{227}--\blpage{239}
(\byear{2003}).
\bcomment{\url{https://doi.org/10.1023/A:1024218716736}}
\end{barticle}
\endbibitem

\bibitem[\protect\citeauthoryear{Vaswani and et~al.}{2017}]{vaswani2017attention}
\begin{bchapter}
\bauthor{\bsnm{Vaswani}, \binits{A.}},
\bauthor{\bsnm{al.}}:
\bctitle{Attention is all you need}.
In: \bbtitle{Advances in Neural Information Processing Systems 30}
(\byear{2017})
\end{bchapter}
\endbibitem

\bibitem[\protect\citeauthoryear{Lan et~al.}{2019}]{lan2019albert}
\begin{botherref}
\oauthor{\bsnm{Lan}, \binits{Z.}}, et al.:
Albert: A lite bert for self-supervised learning of language representations.
arXiv preprint arXiv:1909.11942
(2019).
\url{https://doi.org/10.48550/arXiv.1909.11942}
\end{botherref}
\endbibitem

\bibitem[\protect\citeauthoryear{Devlin et~al.}{2018}]{devlin2018bert}
\begin{botherref}
\oauthor{\bsnm{Devlin}, \binits{J.}}, et al.:
Bert: Pre-training of deep bidirectional transformers for language understanding.
arXiv preprint arXiv:1810.04805
(2018).
\url{https://doi.org/10.48550/arXiv.1810.04805}
\end{botherref}
\endbibitem

\bibitem[\protect\citeauthoryear{Clark et~al.}{2021}]{clark2021canine}
\begin{botherref}
\oauthor{\bsnm{Clark}, \binits{J.H.}},
\oauthor{\bsnm{Garrette}, \binits{D.}},
\oauthor{\bsnm{Turc}, \binits{I.}},
\oauthor{\bsnm{Wieting}, \binits{J.}}:
Canine: Pre-training an efficient tokenization-free encoder for language representation.
arXiv preprint arXiv:2103.06874
(2021).
\url{https://doi.org/10.1162/tacl_a_00448 }
\end{botherref}
\endbibitem

\bibitem[\protect\citeauthoryear{Tan et~al.}{2022}]{tan2022roberta-lstm}
\begin{barticle}
\bauthor{\bsnm{Tan}, \binits{K.L.}}, \betal:
\batitle{Roberta-lstm: A hybrid model for sentiment analysis with transformer and recurrent neural network}.
\bjtitle{IEEE Access}
\bvolume{10},
\bfpage{21517}--\blpage{21525}
(\byear{2022}).
\bcomment{\url{https://doi.org/10.1109/ACCESS.2022.3152828}}
\end{barticle}
\endbibitem

\end{thebibliography}

\end{document}